\documentclass[aps,prd,preprint,nofootinbib,showkeys]{revtex4-2}
\usepackage[utf8]{inputenc}
\usepackage[english]{babel}

\usepackage{graphicx}
\usepackage{tikz}
\usetikzlibrary{quotes,angles,arrows,decorations.markings}

\usepackage{amsmath,amssymb,amsfonts,mathrsfs}
\usepackage{bm,bbm,dsfont,braket,tensor,esint}
\usepackage{cancel,booktabs,multirow,array,makecell}
\newcolumntype{C}{>{\centering\arraybackslash}p{1.8cm}}

\usepackage{color,xcolor}
\usepackage{lipsum}
\usepackage{enumerate}
\usepackage{ulem}
\newif\ifshowchanges
\showchangesfalse   

\ifshowchanges
    \usepackage{changes}
    \setaddedmarkup{\textcolor{red}{#1}}
    \setdeletedmarkup{\textcolor{gray}{\sout{#1}}}
\else
    \usepackage[final]{changes}
\fi

\ifshowchanges

    \newcommand{\redsout}[1]{\textcolor{red}{\sout{\textcolor{black}{#1}}}}
    
\else

    \newcommand{\redsout}[1]{}
\fi

\usepackage{hyperref}
\hypersetup{
    colorlinks,
    breaklinks,
    citecolor=blue,
    filecolor=green,
    linkcolor=blue,
}

\usepackage{slashed}

\definecolor{lime}{HTML}{A6CE39}
\DeclareRobustCommand{\orcidicon}{%
    \begin{tikzpicture}
    \draw[lime, fill=lime] (0,0) 
    circle [radius=0.16] 
    node[white] {{\fontfamily{qag}\selectfont \tiny ID}};
    \draw[white, fill=white] (-0.0625,0.095) 
    circle [radius=0.007];
    \end{tikzpicture}
    \hspace{-2mm}
}

\foreach \x in {A, ..., Z}{%
    \expandafter\xdef\csname orcid\x\endcsname{\noexpand\href{https://orcid.org/\csname orcidauthor\x\endcsname}{\noexpand\orcidicon}} 
}

\begin{document}

\title{Images of Regular Black Holes with Hedgehog Scalar Hair Embedded in Radiatively Inefficient Accretion Flows}

\author{Yuxuan Shi\orcidA{}}
\email{shiyx2280771974@gmail.com}
\affiliation{School of Physics, Beijing Institute of Technology, Beijing 100081, China}

\author{Hao Wei\orcidB{}}
\email{haowei@bit.edu.cn}
\thanks{Corresponding author.}
\affiliation{School of Physics, Beijing Institute of Technology, Beijing 100081, China}

\begin{abstract}
How a black hole's geometric critical scale manifests in an accretion-flow image is governed by photon propagation, plasma radiative processes, and instrumental resolution. In this work, we investigate radiatively inefficient accretion flows (RIAF) surrounding geometrically regular black holes with hedgehog scalar hair. Adopting the mass and distance of $\mathrm{M87}^*$, we generate total-intensity synchrotron images and systematically explore the viewing inclination, disk thickness, and observing frequency. For the fiducial setup at $230\,\mathrm{GHz}$, raising $L/M$ from $0$ to $0.9$ contracts the critical curve diameter by $4.78\%$, shrinks the half-flux radius by $4.07\%$, and lowers the required electron-density scale by $9.15\%$. At $84\,\mathrm{GHz}$, the primary brightness peaks shift farther outside the critical curve due to enhanced synchrotron self-absorption, whereas the sharper emission rings at $230$ and $345\,\mathrm{GHz}$ closely trace the critical curve. Ultimately, we find that although the regular core systematically contracts the emission ring and alters the required plasma density, resolving these deviations from a Schwarzschild black hole demands angular resolution finer than that achievable with current baselines.
\end{abstract}

\maketitle

\newpage
\section{Introduction}
Millimeter-wavelength observations of $\mathrm{M87}^*$ and Sagittarius~$\mathrm{A}^*$ by the Event Horizon Telescope (EHT) have revealed bright emission rings enclosing a central brightness depression~\cite{akiyama2019firsta,akiyama2022firsta}, which theoretical modeling links to gravitational lensing, hot accretion flows, and the black hole mass-to-distance ratio~\cite{akiyama2019firstb,akiyama2019firstc,akiyama2022firstb,akiyama2022firstc}. Testing a specific spacetime metric with such images requires determining how metric-induced shifts in geometric scale survive plasma emission, absorption, and finite resolution. While the photon capture boundary defines a theoretical critical curve extensively characterized across spherically symmetric spacetimes~\cite{synge1966the,falcke2000viewing,perlick2022calculating,shaikh2019shadows,wang2026extended}, a realistic image mixed direct emission, lensed photon rings, and higher-order winding rays weighted by the emission profile~\cite{gralla2019black,gralla2020lensing,narayan2019the,yang2026observational}. Although the narrow ring maintains a relation to the geometric shadow in optically thin accretion flows~\cite{ozel2022black,younsi2023black}, the disk thickness, fluid kinematics, and emission morphology can shift ring widths and peak radii~\cite{vincent2022images,urso2025gravity,zeng2026impact}, motivating an evaluation of both geometric and radiation scales.

RIAF provides the physical foundation for modeling low luminosity black holes~\cite{narayan1994advection,narayan1995advection,yuan2014hot}, where synchrotron emission and absorption are governed by electron thermodynamics and magnetic fields~\cite{yuan2003nonthermal,ressler2015electron,chael2018the,noble2007simulating,moscibrodzka2016general,zhao2026signatures}, with being modulated by Doppler shifts from radial infall and rotation~\cite{pu2016the,pu2018probing}. Parameterized prescriptions depict these effects under appropriate assumptions. Meanwhile observing frequency and flux calibration fix the compared plasma states. Notably, the larger M87$^*$ ring at $86\,\mathrm{GHz}$ relative to $230\,\mathrm{GHz}$~\cite{lu2023a}, considering the studies of frequency-dependent centroids and self-absorbed multifrequency rings~\cite{broderick2006frequency,desire2025multifrequency}, emphasizes the impact of optical depth and emitting-region extent. Moreover, matching different models to an identical total flux requires rescaling the electron density. When the magnetic field scales with plasma pressure, this alters the field strength, emissivity, and absorption concurrently, linking comparisons to the basic plasma state~\cite{akiyama2019firstb,akiyama2022firstb}.

Black holes with regular cores~\cite{dymnikova1992vacuum,hayward2006formation,fan2016construction,lan2023regular} and scalar hair~\cite{herdeiro2014kerr,herdeiro2015asymptotically} offer the specific circumstance to probe additional fundamental length scales, as metric deformations modify both photon orbits and the emitting region~\cite{gan2021photon,meng2023images,saurabh2024imaging}. Recently, Bahamonde constructed an exact geometrically regular black hole with hedgehog scalar hair in Einstein gravity coupled to a fixed-norm scalar triplet and an auxiliary three-form sector~\cite{bahamonde2026geometrically}. In the cubic kinetic regular branch, action parameters fix the core scale $L = \eta/\mu_\star$ controlling the scalar energy-density profile. Although asymptotic deviations from the Schwarzschild spacetime occur at $\mathcal{O}(r^{-4})$, the mass distribution varies appreciably with the parameter $L/M$, which provide a framework to study how modifications change the cumulative-flux radii and plasma densities~\cite{bahamonde2026geometrically}.

Finite instrumental resolution further determines whether these radial shifts remain detectable in two-dimensional black hole images. Building on radiative-transfer studies of non-Kerr hot flows~\cite{mizuno2018the,younsi2023black,jiang2024shadows,yin2026bright,yang2026shadow} and high-resolution photon-ring or inner-shadow diagnostics~\cite{johnson2020universal,chael2021observing,zhao2026signatures,yang2026shadow}, we adopt the mass and distance of $\mathrm{M87}^*$ to study RIAF images for different values for $L/M$. Varying the viewing inclination, flow thickness, and frequency at a fixed flux, we compare the critical scale against cumulative-flux radii using density calibrations, optical depths, and intensity cross-sections. We then evaluate the cross-correlation mismatch and decomposed structural dissimilarity to quantify morphological variation induced by the core surviving at finite resolution.

The remainder of this paper is organized as follows. Section~\ref{sec2} introduces the spacetime about the regular black-hole with hedgehog scalar hair. Section~\ref{sec3} details the RIAF model and methodology. Section~\ref{sec4} simulates the inclination, thickness, and frequency sequences, while Section~\ref{sec5} analyzes finite-resolution effects and image-comparison metrics. Finally, Section~\ref{sec6} discusses our findings and summarizes our conclusions.


\section{Review of the regular black hole with hedgehog scalar hair}
\label{sec2}
In Bahamonde's model, an auxiliary three-form and a fixed-modulus scalar triplet are coupled to Einstein gravity~\cite{bahamonde2026geometrically,nashed2026odd}. With the hedgehog configuration $\Phi^I=\eta n^I(\theta,\varphi)$ and $n^In^I=1$, the triplet satisfies $\delta_{IJ}\Phi^I\Phi^J=\eta^2$. The model is
\begin{align}
S=\int\mathrm{d}^4x\sqrt{-g}\left[\dfrac{R}{16\pi G}-\mathcal{C}\widehat{K}(Y)+\lambda(\delta_{IJ}\Phi^I\Phi^J-\eta^2)-\dfrac{1}{3!}\dfrac{\varepsilon^{\mu\nu\rho\sigma}}{\sqrt{-g}}\mathcal{A}_{\mu\nu\rho}\nabla_\sigma\mathcal{C}\right],
\end{align}
where
\begin{align}
Y=\dfrac12\delta_{IJ}\nabla_\mu\Phi^I\nabla^\mu\Phi^J,
\end{align}
and an arbitrary function of the kinetic term
\begin{align}
\widehat{K}(Y)=\left(\dfrac{Y}{Y+\mu_\star^2}\right)^3.
\end{align}
For an isolated background, $\mathcal C=\rho_0$ can be written. The auxiliary-field equation yields $\nabla_\mu\mathcal C=0$. Since $Y=\eta^2/r^2$ in the hedgehog shape, the core scale is
\begin{align}
L=\dfrac{\eta}{\mu_\star},
\end{align}
and the density supporting the geometry is
\begin{align}
\rho(r)=\dfrac{\rho_0}{\left[1+\left(r^2/L^2\right)\right]^3}.
\end{align}
In this case, we set $c=1$. The core density of the background scalar source is $\rho_0$, and its radial scale is determined by the parameter $L$, which is determined by the action parameters $\eta$ and $\mu_\star$. An isolated solution's mass fluctuates within a fixed theory via the integration constant $\rho_0$. The spherically symmetric line element is
\begin{align}
\mathrm{d}s^2=-f(r)\mathrm{d}t^2+\dfrac{\mathrm{d}r^2}{f(r)}
+r^2\left(\mathrm{d}\theta^2+\sin^2\theta\mathrm{d}\varphi^2\right),
\end{align}
where
\begin{align}
\label{eq:fr}
f(r)=1-\dfrac{4M}{\pi r}\left[\arctan\left(\dfrac{r}{L}\right)+\dfrac{r}{L}\dfrac{r^2/L^2-1}{\left(1+r^2/L^2\right)^2}\right].
\end{align}
We used the gravitational radius $r_g=GM/c^2$ to convert dimensionless quantities into physical angular scales. Adopting the mass of $M=6.4\times10^9\,M_\odot$ and the distance of $D = 16.9\,\mathrm{Mpc}$, we obtain the angular gravitational radius of $\theta_g = r_g/D \approx 3.7381\,\mu\mathrm{as}$~\cite{chen2026images}. In geometric units $G = c = 1$, the dimensionless parameter choices $L/M = 0.3, 0.6, 0.9$, respectively, with the Schwarzschild spacetime serving as the $L/M = 0$ reference case~\cite{psaltis2020gravitational,kocherlakota2021constraints}. Furthermore, a critical value $L/M\approx1.0237$ could be yielded by the conditions
\begin{align}
f(r_h)=0,\qquad \left.\dfrac{\mathrm{d}f}{\mathrm{d}r}\right|_{r_h}=0.
\end{align}
ensuring that all three considered cases possess both inner and outer horizons \cite{bahamonde2026geometrically}.

\begin{figure}[tbp]
\centering
\includegraphics[width=\textwidth]{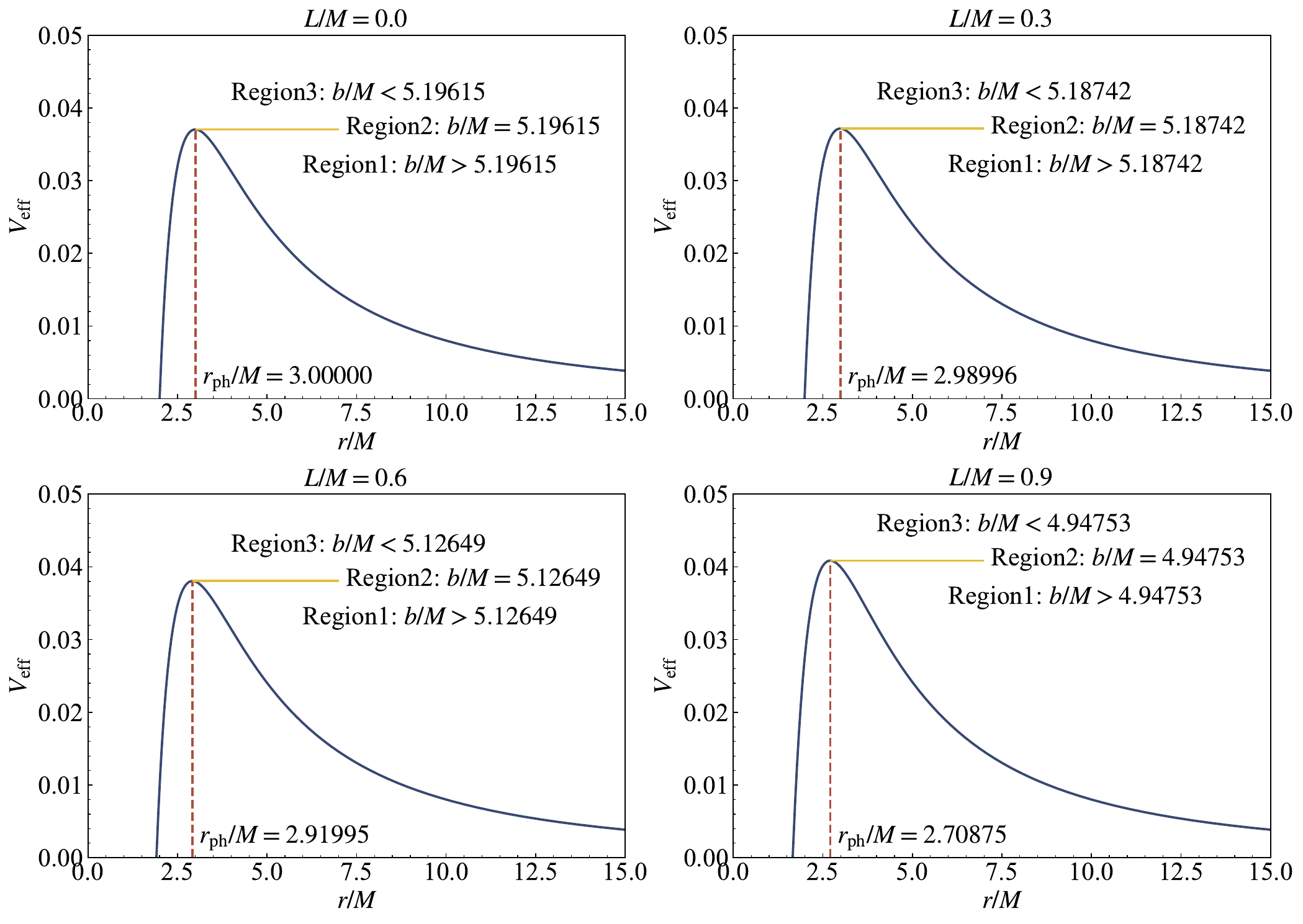}
\caption{Effective photonic potential $V_{\mathrm{ph}}(r)$ as a function of the dimensionless radius $r/M$ for the regular black hole with hedgehog scalar hair. The peak of each curve marks the photon sphere radius $r_{\mathrm{ph}}$, which defines the critical impact parameter $b_p$. Depending on $b$, photon trajectories fall into three distinct regimes: scattering back to infinity (Region 1, $b > b_p$), the critical unstable circular orbit (Region 2, $b = b_p$), and capture by the black hole (Region 3, $b < b_p$).}
\label{fig:Vph}
\end{figure}

In the spacetime described by Eq.~\eqref{eq:fr}, we consider photon motion restricted to the equatorial plane, $\theta = \pi/2$, $\dot{\theta} = 0$. The corresponding Lagrangian is given by
\begin{align}
\mathcal{L} &= \dfrac{1}{2} g_{\mu\nu} \dot{x}^{\mu} \dot{x}^{\nu} \\\notag
&= \dfrac{1}{2} \left[ -f(r)\dot{t}^2 + \dfrac{\dot{r}^2}{f(r)} + r^2\dot{\varphi}^2 \right],
\end{align}
where the overdot denotes differentiation with respect to the affine parameter $\lambda$, i.e., $\dot{x}^{\mu} = \mathrm{d}x^{\mu}/\mathrm{d}\lambda$. Since the metric components are independent of the coordinates $t$ and $\varphi$, the photon's energy $E$ and angular momentum $L_z$ are conserved:
\begin{equation}
\label{eq:E+L}
-E=\dfrac{\partial\mathcal{L}}{\partial\dot{t}}=f(r)\dot{t}, \qquad
L_z=\dfrac{\partial\mathcal{L}}{\partial\dot{\varphi}}=r^2\dot{\varphi}.
\end{equation}
Imposing the null condition $g_{\mu\nu}\dot{x}^{\mu}\dot{x}^{\nu}=0$, the radial equation of photon takes the form
\begin{equation}
\dot{r}^2+V_{\mathrm{eff}}(r)=0,
\end{equation}
with the effective potential given explicitly by
\begin{equation}
V_{\mathrm{eff}}(r)=\dfrac{L_z^2}{r^2}\left\{1-\dfrac{4M}{\pi r}\left[\arctan\left(\dfrac{r}{L}\right)+\dfrac{r}{L}\dfrac{r^2/L^2-1}{\left(1+r^2/L^2\right)^2}\right]\right\}-E^2.
\end{equation}
To characterize photon trajectories in terms of the impact parameter $b \equiv |L_z|/E$, it is also convenient to rewrite the radial equation as 
\begin{align}
\dfrac{\dot{r}^2}{L_z^2}+V_{\mathrm{ph}}(r)=\dfrac{1}{b^2},
\end{align}
where $V_{\mathrm{ph}}(r) \equiv f(r)/r^2$ denotes the reduced photonic potential shown in Fig.~\ref{fig:Vph}, which illustrates the function $V_{\mathrm{ph}}(r)$ for various values of $L/M$. The potential vanishes at the outer event horizon $r_h$, where infalling light rays terminate; accordingly, we set $r_h$ as the termination radius for ray integration in our subsequent imaging calculations. Starting from $r_h$, $V_{\mathrm{ph}}(r)$ increases monotonically, attains a global maximum at the photon sphere radius $r_{\mathrm{ph}}$, and subsequently decays asymptotically to zero as $r \to \infty$. Thus, $V_{\mathrm{ph}}(r)$ naturally forms a potential barrier for ingoing rays.

Unstable circular null geodesics around the black hole form the photon sphere, where orbits remain at a constant radius $r = r_{\mathrm{ph}}$ satisfying $\dot{r} = \ddot{r} = 0$. In terms of the effective potential, circular photon orbits are determined by the conditions
\begin{align}
V_{\mathrm{eff}}(r_{\mathrm{ph}})=\left.\dfrac{\partial V_{\mathrm{eff}}(r)}{\partial r}\right|_{r = r_{\mathrm{ph}}}=0,
\end{align}
which immediately yield the critical impact parameter~\cite{bozza2002gravitational,perlick2022calculating}
\begin{align}
b_p=\dfrac{r_{\mathrm{ph}}}{\sqrt{f(r_{\mathrm{ph}})}}=\dfrac{1}{\sqrt{V_{\mathrm{ph}}(r_{\mathrm{ph}})}}.
\end{align}
Radial instability of the photon sphere further requires that the potential attains a local maximum at $r = r_{\mathrm{ph}}$~\cite{cardoso2009geodesic}, so that
\begin{equation}
\left.\dfrac{\partial^2 V_{\mathrm{eff}}(r)}{\partial r^2}\right|_{r_{\mathrm{ph}}}<0.
\end{equation}

As depicted in Fig.~\ref{fig:Vph}, ingoing rays with $b > b_p$ (Region~1) cannot overcome the barrier and are reflected back at the radial turning point. Conversely, in Region~3 ($b < b_p$), ingoing photons inevitably plunge into the black hole. Specifically, in the critical case $b = b_p$ (Region~2), light rays approach the photon sphere and orbit the black hole indefinitely~\cite{synge1966the,gralla2019black}.


\section{Radiatively Inefficient Accretion Flow}
\label{sec3}
For primary EHT observational targets such as $\mathrm{M87}^*$ and $\mathrm{Sgr~A}^*$, the surrounding hot, magnetized plasma can be described phenomenologically by the RIAF~\cite{narayan1994advection,narayan1995advection,yuan2014hot,akiyama2019firstb,akiyama2022firstb,yang2026shadow}. In this work, we model the emitting plasma as a stationary, axisymmetric parameterized RIAF suitable for black holes, where millimeter-wavelength emission is dominated by thermal synchrotron radiation and self-absorption~\cite{pu2016the,pu2018probing,yin2026bright,chen2026images}. Within this semi-analytic prescription, the spatial distributions of the electron number density $n_e$ and electron temperature $T_e$ are written as~\cite{broderick2006frequency,pu2016the}
\begin{align}
\label{eq:ne}
n_e(r,\theta)&=n_{e,0}\left(\dfrac{r}{r_g}\right)^{-\delta}
\exp\left(-\dfrac{\cot^2\theta}{2H^2}\right),\\
\label{eq:Te}
T_e(r)&=T_{e,0}\left(\dfrac{r}{r_g}\right)^{-\alpha},
\end{align}
where $r_g=GM/c^2$ denotes the gravitational radius. Near the equatorial plane, the dimensionless parameter $H$ characterizes the thinkness of the disk. Here, $n_{e,0}$ and $T_{e,0}$ represent the reference electron number density and temperature normalizations, respectively, with radial power-law indices $\delta$ and $\alpha$ constrained by broadband spectral energy distributions and very-long-baseline interferometry (VLBI) brightness temperature measurements~\cite{mahadevan1997scaling,yuan2003nonthermal,broderick2006frequency}. Assuming the magnetic pressure is a fixed fraction of the local ion virial pressure, the magnetic field strength $B$ satisfies~\cite{pu2016the,chen2026images}
\begin{align}
\label{eq:magnetic}
\dfrac{B^2}{8\pi}=\dfrac{n_em_pc^2}{60r},
\end{align}
where $m_p$ is the proton mass. At any fixed radius, the magnetic field scales with the electron density normalization as $B \propto n_{e,0}^{1/2}$. To capture realistic accretion kinematics, the fluid four-velocity $u^\mu$ is constructed by interpolating between a Keplerian rotating branch and a radially infalling free-fall branch~\cite{pu2016the,pu2018probing}. Outside the innermost stable circular orbit (ISCO), the Keplerian branch satisfies $u_K^r = 0$ and $\Omega_K = \sqrt{f'(r)/(2r)}$. Inside the ISCO, the flow plunges inward while conserving the specific energy and angular momentum evaluated at the ISCO boundary along the same latitude. Combining both regimes, the four-velocity takes the form
\begin{align}
u^{\mu}=\left(u^t,u^r,0,\Omega u^t\right),
\end{align}
with the radial velocity and angular velocity prescribed by
\begin{align}
u^r&=u_K^r+\kappa_{\mathrm{ff}}(u_{\mathrm{ff}}^r-u_K^r),\\
\Omega&=\kappa_K\Omega_K.
\end{align}
Here, $\left(u^r_K, \Omega_K\right)$ and $\left(u^r_{\mathrm{ff}}, \Omega_{\mathrm{ff}}\right)$ denote the radial four-velocities and angular velocities of Keplerian and free-fall branches, respectively. Their relative weights are governed by the interpolation parameters $\kappa_{\mathrm{ff}}$ and $\kappa_K$, which recover pure Keplerian rotation for $\left(\kappa_{\mathrm{ff}}, \kappa_K\right) = (0, 1)$ and pure radial free fall for $(1, 0)$~\cite{pu2016the}. Due to the accretion flow onto $\mathrm{M87}^*$ being expected to exhibit both radial advection and sub-Keplerian rotation~\cite{yuan2014hot,akiyama2019firstb}, we adopt an equal-weight hybrid prescription with $\kappa_{\mathrm{ff}} = \kappa_K = 0.5$. With vanishing poloidal velocity ($u^\theta = 0$) and $u^\varphi = \Omega u^t$, imposing the timelike normalization condition $u_\mu u^\mu = -1$ determines the temporal component as
\begin{align}
u^t=\left[\dfrac{1+\left(u^r\right)^2/f(r)}{f(r)-r^2\sin^2\theta\,\Omega^2}\right]^{1/2}.
\end{align}

At the primary EHT observing frequency of $230\,\mathrm{GHz}$, millimeter emission from the inner accretion flow is predominantly governed by thermal synchrotron processes, allowing non-thermal contributions to be safely neglected~\cite{mahadevan1996harmony,yuan2003nonthermal,akiyama2019firstb}. We therefore compute the synchrotron radiative coefficients assuming the Maxwell--J\"{u}ttner electron distribution~\cite{leung2011numerical,dexter2016a,pandya2016polarized}. To match the observed millimeter flux density of $\mathrm{M87}^*$, the characteristic electron density and temperature scales are calibrated to $n_{e,0} \approx 10^6\,\mathrm{cm}^{-3}$ and $T_{e,0} \approx 10^{11}\,\mathrm{K}$~\cite{akiyama2019firstb,chen2026images}. All model parameters are summarized in Tab.~\ref{tab:params}.

\begin{table}[tp]
\centering
\caption{Summary of the fiducial physical and observational parameters adopted for $\mathrm{M87}^*$.}
\label{tab:params}
\setlength{\tabcolsep}{14pt}
\begin{tabular}{lll}
\toprule
Symbol & Fiducial Value & Physical Meaning \\
\midrule
$M_{\mathrm{BH}}$ & $6.4 \times 10^9\,M_\odot$ & Mass of the central black hole \\
$D_s$ & $16.9 \times 10^6\,\mathrm{pc}$ & Source distance \\
$\delta$ & $1.1$ & Power-law slope of electron density $n_e$ \\
$\alpha$ & $0.84$ & Power-law slope of electron temperature $T_e$ \\
$\kappa_K$ & $0.5$ & Keplerian rotation weight \\
$\kappa_{\mathrm{ff}}$ & $0.5$ & Radial free-fall weight \\
$n_{e,0}$ & $\approx 10^6\,\mathrm{cm}^{-3}$ & Reference electron number density \\
$T_{e,0}$ & $\approx 10^{11}\,\mathrm{K}$ & Reference electron temperature \\
\bottomrule
\end{tabular}
\end{table}

We employ a backward ray-tracing and general relativistic polarized radiative transfer (GRRT) scheme to generate synthetic images~\cite{dexter2009a,vincent2011gyoto,younsi2012general,dexter2016a,moscibrodzka2018ipole,gold2020verification,prather2023comparison}. Because the null geodesics depend solely on the background metric, trajectories integrated for a given core scale $L/M$ and inclination $i$ remain invariant across different plasma thicknesses and observing frequencies. Along each ray, the frequency redshift factor $g$ relating the observed frequency $\nu_o$ to the local comoving fluid-frame frequency $\nu_e$ is given by
\begin{align}
\label{eq:ratio}
g=\dfrac{\nu_o}{\nu_e}
=\dfrac{-k_\mu u_o^\mu}{-k_\mu u_e^\mu},
\end{align}
which contains both gravitational redshift and relativistic Doppler boosting. The integration utilizes Lorentz-invariant quantities (such as $I_\nu/\nu^3$) and parallel-transports the local polarization basis along the geodesic~\cite{dexter2016a,moscibrodzka2018ipole}. Although our primary analysis focuses on the emergent total intensity $I_\nu$, we evolve the full Stokes vector $\boldsymbol S_\nu=\left(I_\nu,Q_\nu,U_\nu,V_\nu\right)^{\mathsf{T}}$ to self-consistently capture polarized emission, absorption, and Faraday mixing. In the local fluid rest frame, the polarized radiative transfer equation reads~\cite{shcherbakov2008propagation,dexter2016a,pandya2016polarized}
\begin{align}
\label{eq:dSds}
\dfrac{\mathrm{d}\boldsymbol S_\nu}{\mathrm{d}s}
=\boldsymbol j_\nu-\boldsymbol K_\nu\boldsymbol S_\nu,
\end{align}
where $\boldsymbol j_\nu = \left(j_I, j_Q, j_U, j_V\right)^{\mathsf{T}}$ is the emissivity vector and the propagation matrix $\boldsymbol K_\nu$ takes the form
\begin{align}
\boldsymbol K_\nu=
\begin{pmatrix}
\alpha_I&\alpha_Q&\alpha_U&\alpha_V\\
\alpha_Q&\alpha_I&\rho_V&-\rho_U\\
\alpha_U&-\rho_V&\alpha_I&\rho_Q\\
\alpha_V&\rho_U&-\rho_Q&\alpha_I
\end{pmatrix}.
\end{align}
Here, the thermal synchrotron emissivities $j_{I,Q,U,V}$ are evaluated using the analytical fitting formulae of Dexter~\cite{dexter2016a}, the absorption coefficients $\alpha_{I,Q,U,V}$ follow from Kirchoff's law of thermal equilibrium, and conversion coefficients $\rho_{Q,U,V}$ are computed via relativistic thermal plasma prescriptions~\cite{shcherbakov2008propagation,dexter2016a,pandya2016polarized}.

To resolve the fine structure of narrow photon rings along the horizontal cross-sections ($Y=0$), we employ locally refined sub-pixel sampling. Integrating the emergent specific intensity $I_\nu$ (in cgs units of $\mathrm{erg\,s^{-1}\,cm^{-2}\,Hz^{-1}\,sr^{-1}}$) over the observer's screen yields the total flux density
\begin{align}
F_\nu=\int I_\nu\,\mathrm{d}\Omega\simeq10^{23}\sum_{i,j}I_{\nu,ij}(\Delta\vartheta)^2\,\mathrm{Jy},
\end{align}
where $I_{\nu,ij}$ is the specific intensity at pixel $(i,j)$, $\Delta\vartheta$ is the pixel angular spacing in radians, and the factor $10^{23}$ converts the cgs flux to Jansky. Based on $F_\nu$, we characterize the radial emission distribution via the cumulative-flux radii $R_{50}$ and $R_{90}$, defined as the radii of concentric circles centered on the projected black-hole origin that enclose $50\%$ and $90\%$ of the total flux, respectively; a smaller $R_{50}$ reflects a more compact central ring, whereas $R_{90}$ traces the outer extended emission. To quantify Doppler-induced azimuthal asymmetry, we evaluate both the left-half flux fraction, 
\begin{align}
f_{\rm left}\equiv\dfrac{F_\nu(X<0)}{F_\nu},
\end{align}
and the intensity-weighted horizontal centroid,
\begin{align}
\overline{X}\equiv\dfrac{1}{F_\nu}\int X I_\nu\,\mathrm{d}\Omega,
\end{align}
measured relative to the projected black-hole center. Together, $\overline{X}$, $f_{\rm left}$, and $R_{50}$ provide complementary diagnostics of the horizontal centroid shift, left--right flux partition, and radial compactness of the image.


\section{Imaging Responses at Fixed Total Flux Density}
\label{sec4}

\subsection{Viewing Inclination}
\begin{figure}[tbp]
\centering
\includegraphics[width=\textwidth]{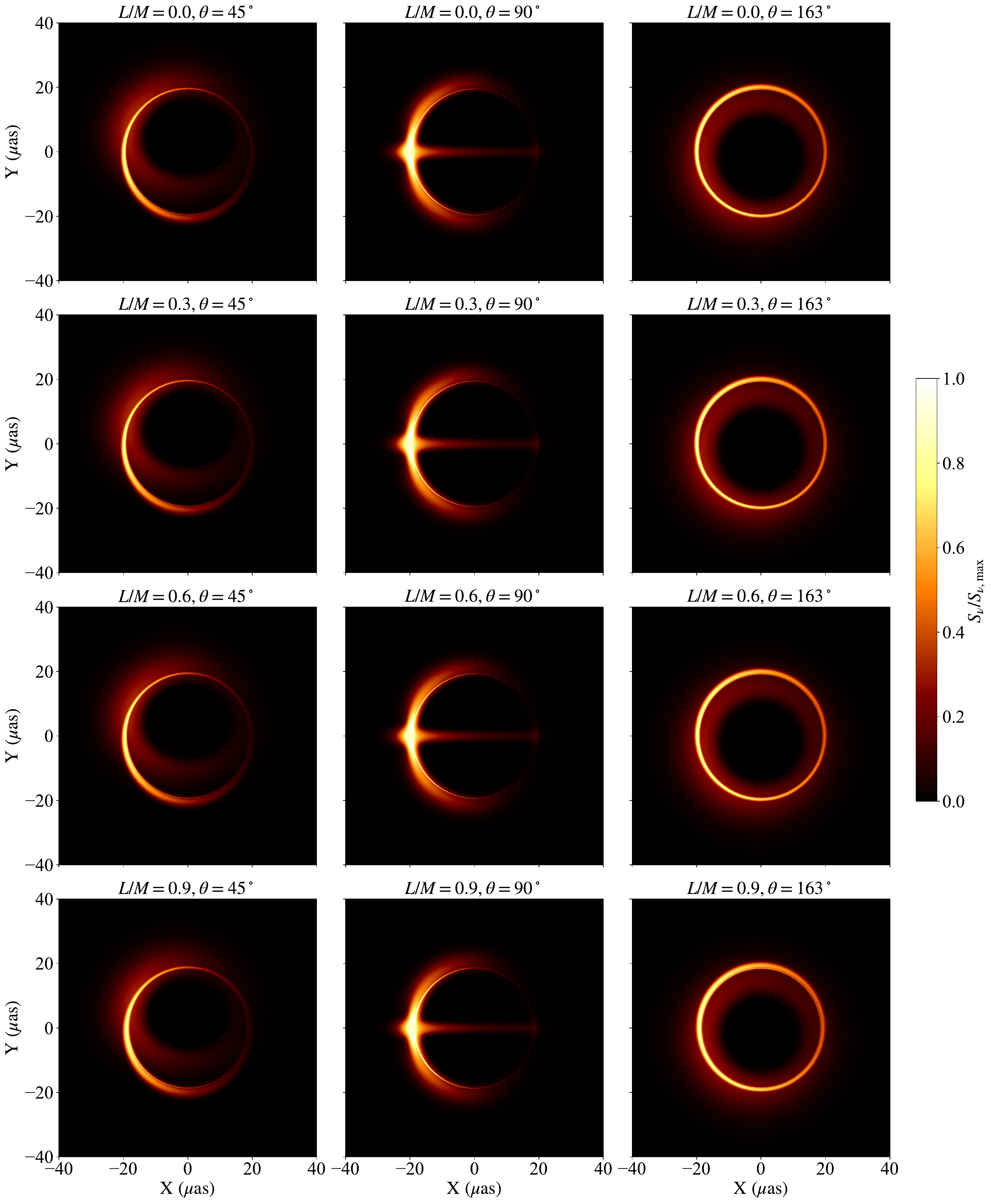}
\caption{Total-intensity synchrotron images across different viewing inclinations. Rows from top to bottom correspond to core scales $L/M=0, 0.3, 0.6, 0.9$, while columns from left to right correspond to viewing inclinations $\theta=45^\circ, 90^\circ, 163^\circ$ at fixed $\nu_o=230\,\mathrm{GHz}$ and $H=0.1$.}
\label{fig:Inclination_1}
\end{figure}

The viewing inclination $\theta$ strongly modulates the azimuthal brightness distribution of the accretion flow by altering the line-of-sight velocity projection, the geometric path length through the plasma, and the pitch angle $\vartheta_B$ between the photon wavevector and the toroidal magnetic field~\cite{pu2016the,pu2018probing,akiyama2019firstb}. Fig.~\ref{fig:Inclination_1} compares the total-intensity images at $\theta=45^\circ, 90^\circ, 163^\circ$ for $\nu_{o}=230\,\mathrm{GHz}$ and $H=0.1$. In view of the spherically symmetric background spacetime, the geometric critical curve at a given core scale $L/M$ remains invariant across all viewing angles. However, the comoving frequency factor $-k_\mu u_e^\mu$ in Eq.~\eqref{eq:ratio} varies systematically with the projected fluid four-velocity along the line of sight, while the magnetic pitch angle enters the local synchrotron emissivity and absorption coefficients. Consequently, the observed brightness arcs reflect the combined imprint of relativistic Doppler boosting, directional synchrotron emission, and path-integrated self-absorption.

Among the three inclinations, the edge-on equatorial view ($\theta=90^\circ$) exhibits the most pronounced left--right flux asymmetry. In the Schwarzschild case ($L/M=0$), the left-half flux fraction $f_{\rm left}$ reaches $88.39\%$ at $\theta=90^\circ$, compared to $79.22\%$ at $\theta=45^\circ$ and $62.38\%$ at the nearly face-on $\mathrm{M87}^*$ orientation of $\theta=163^\circ$~\cite{akiyama2019firsta,akiyama2019firstb}, with horizontal flux centroids located at $\overline X = -12.626, -9.744, -4.247\,\mu\mathrm{as}$, respectively. At $\theta=90^\circ$, light rays propagate directly along the dense equatorial disk plane where the azimuthal fluid velocity has its maximal line-of-sight projection, producing both a strongly beamed approaching (left) crescent and a prominent horizontal emission lane across the central depression. Conversely, at $\theta=163^\circ$, the reduced line-of-sight velocity projection yields a much more symmetric, continuous emission ring enclosing a deep central brightness depression~\cite{gralla2019black,narayan2019the}.

Increasing the regular core scale $L/M$ systematically contracts the radial emission structure across all inclinations. Specifically, raising $L/M$ from $0$ to $0.9$ reduces the half-flux radius $R_{50}$ from $20.440$ to $19.568\,\mu\mathrm{as}$ ($\Delta R_{50} = -0.872\,\mu\mathrm{as}$) at $\theta=45^\circ$, from $20.752$ to $19.904\,\mu\mathrm{as}$ ($\Delta R_{50} = -0.848\,\mu\mathrm{as}$) at $\theta=90^\circ$, and from $20.205$ to $19.382\,\mu\mathrm{as}$ ($\Delta R_{50} = -0.823\,\mu\mathrm{as}$, or $4.07\%$) at $\theta=163^\circ$. Concurrently, the horizontal flux centroids shift slightly inward to $\overline X = -9.353, -12.304, -4.038\,\mu\mathrm{as}$ at $\theta=45^\circ, 90^\circ, 163^\circ$, respectively. This clean separation indicates that within fixed-flux RIAF models, the radial cumulative-flux scale responds robustly to the regular core geometry, whereas the global left--right asymmetry is dictated almost entirely by the viewing geometry.

\begin{figure}[tbp]
\centering
\includegraphics[width=\textwidth]{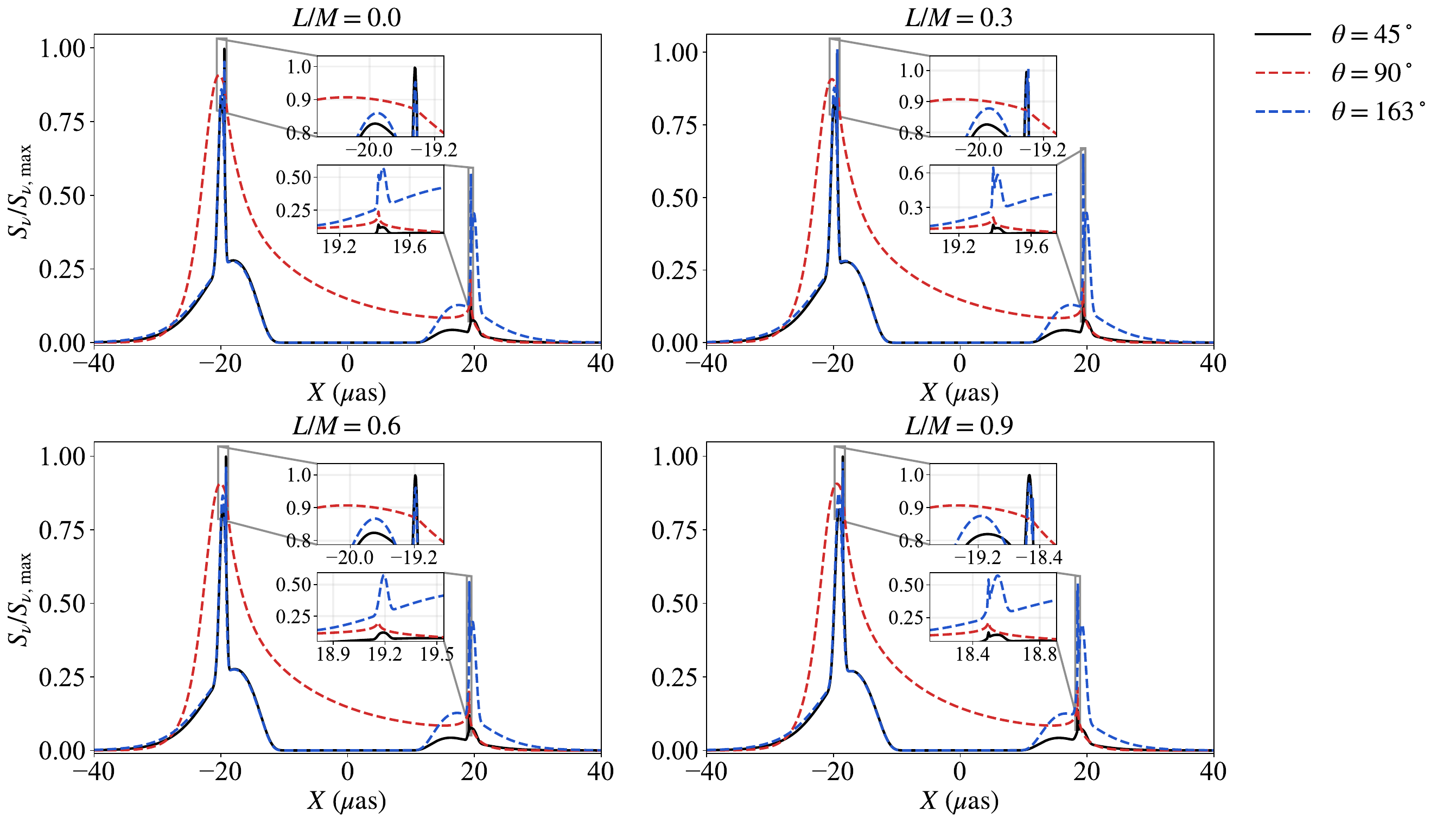}
\caption{Horizontal intensity cross-sections along $Y=0$ regrouped by fixed core scale $L/M$ across viewing inclinations $\theta=45^\circ, 90^\circ, 163^\circ$.}
\label{fig:Inclination_2-1}
\end{figure}

\begin{figure}[tbp]
\centering
\includegraphics[width=\textwidth]{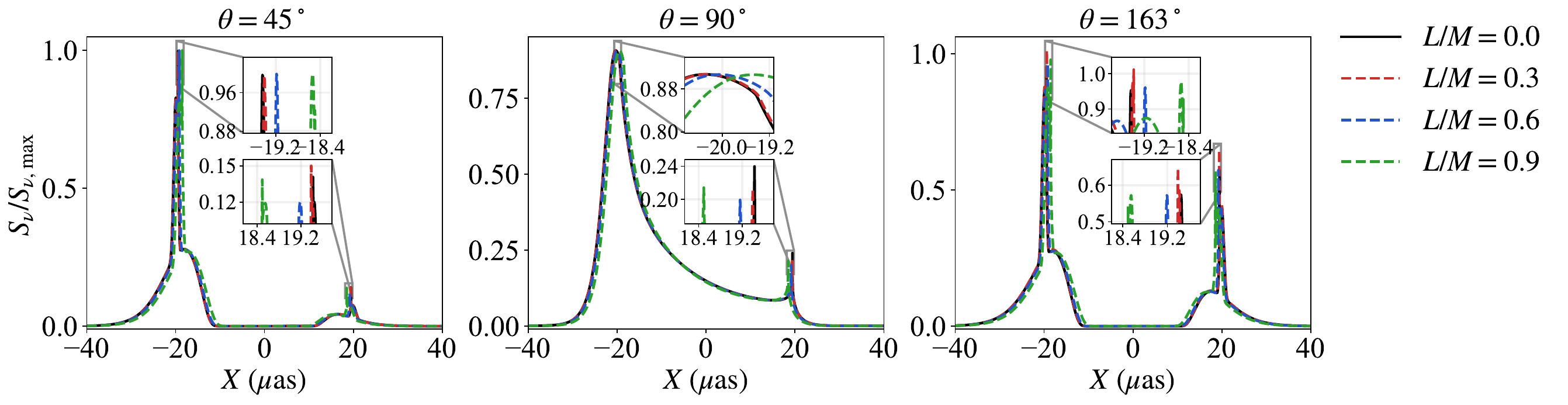}
\caption{Horizontal intensity cross-sections along $Y=0$ for different values of inclination $\theta=45^\circ, 90^\circ, 163^\circ$, comparing the four core scales $L/M = 0, 0.3, 0.6, 0.9$.}
\label{fig:Inclination_2-2}
\end{figure}

The one-dimensional horizontal intensity cross-sections along $Y=0$ is displayed in Figs.~\ref{fig:Inclination_2-1} and \ref{fig:Inclination_2-2}. As the parameter $L/M$ increases, the sharp photon-ring peaks in the low-inclination view ,$\theta=163^\circ$, shift inward in lockstep with the contracting geometric critical curve, whereas at $\theta=90^\circ$ the main intensity peak blends lensed ring emission with broad foreground disk emission along the equatorial sightline. The horizontal profiles thus capture the net radiative equilibrium after accounting for both metric contraction and the accompanying plasma density and magnetic-field rescaling.


\subsection{Disk Thinkness}
\begin{figure}[tbp]
\centering
\includegraphics[width=\textwidth]{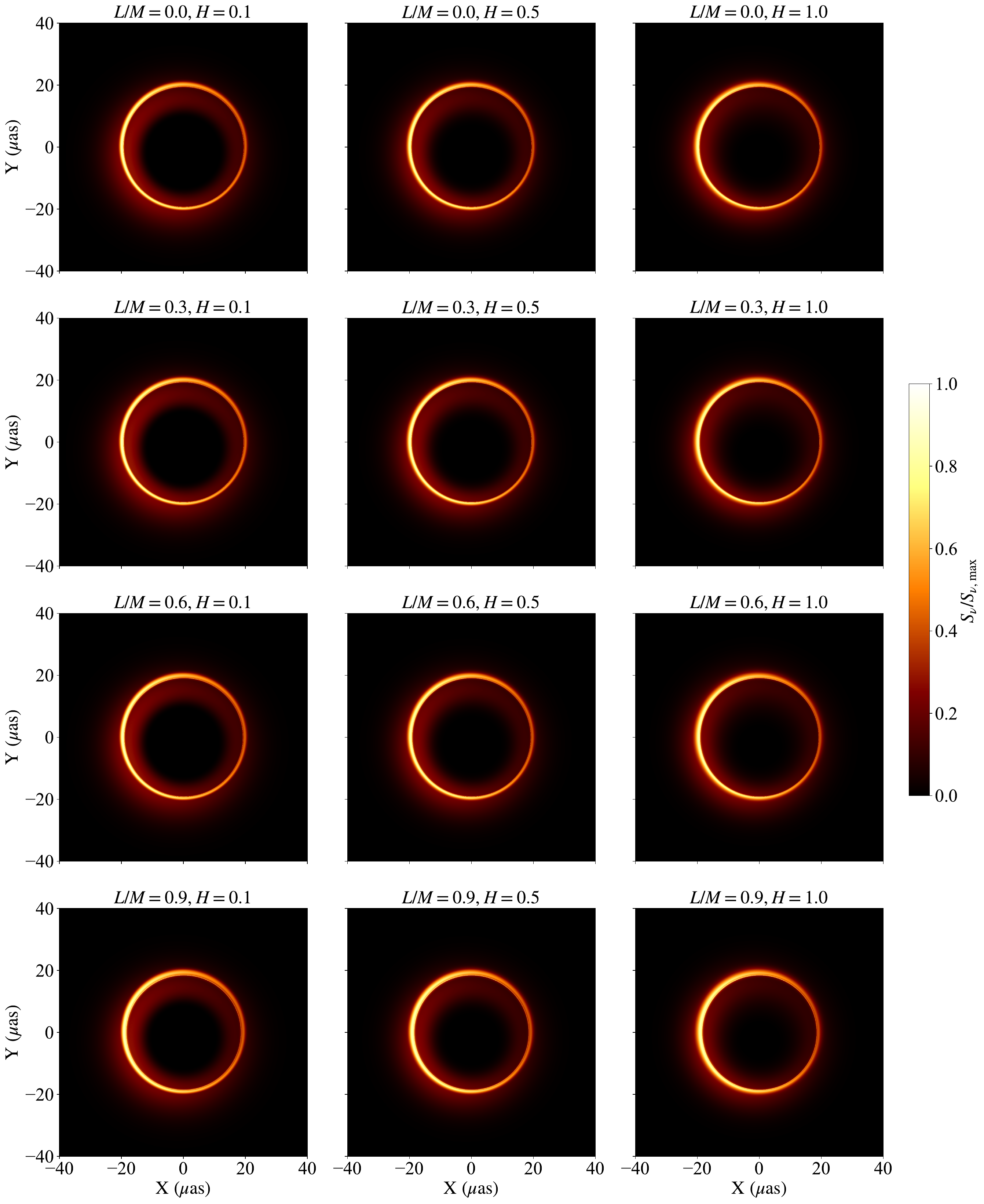}
\caption{Total-intensity synchrotron images across different disk thicknesses. Columns from left to right correspond to $H=0.1, 0.5, 1.0$, and rows from top to bottom correspond to $L/M=0, 0.3, 0.6, 0.9$, evaluated at $\theta=163^\circ$ and $\nu_o=230\,\mathrm{GHz}$.}
\label{fig:RIAF_thickness_1}
\end{figure}

The thickness $H$ governs the vertical extent of the electron density profile via Eq.~\eqref{eq:ne}, thereby modulating the line-of-sight column densities along the rays and the resulting synchrotron emission and self-absorption. For each value of $H$, the electron density normalization $n_{e,0}$ is independently recalibrated to maintain a fixed total flux density of $0.5\,\mathrm{Jy}$ following the procedure in Ref.~\cite{vincent2022images}. Fig.~\ref{fig:RIAF_thickness_1} presents the synthetic images for $H=0.1, 0.5, 1.0$ at the fiducial viewing angle $\theta=163^\circ$ and $\nu_o=230\,\mathrm{GHz}$. As indicated by Eq.~\eqref{eq:ne}, increasing $H$ weakens the attenuation $\exp\left[-\cot^2\theta/\left(2H^2\right)\right]$ away from the equatorial plane and expands the emitting volume into higher latitudes. Consequently, at fixed $L/M=0$, matching the $0.5\,\mathrm{Jy}$ target flux as the flow thickens from $H=0.1$ to $0.5$ and $1.0$ reduces the required density normalization from $n_{e,0} = 1.2011\times 10^6\,\mathrm{cm}^{-3}$ to $0.6685\times 10^6\,\mathrm{cm}^{-3}$ and $0.5412\times 10^6\,\mathrm{cm}^{-3}$, respectively. Similarly, across the core parameter $L/M$, $n_{e,0}$ declines monotonically from $0.6685$ to $0.6022$ (in $10^6\,\mathrm{cm}^{-3}$) for $H=0.5$ and from $0.5412$ to $0.4861$ for $H=1.0$. Eq.~\eqref{eq:magnetic} implies that the lower density normalization reduces the equatorial magnetic field strength via $B\propto n_{e,0}^{1/2}$ while redistributing synchrotron emissivity and absorption across latitudes.

\begin{figure}[tbp]
\centering
\includegraphics[width=\textwidth]{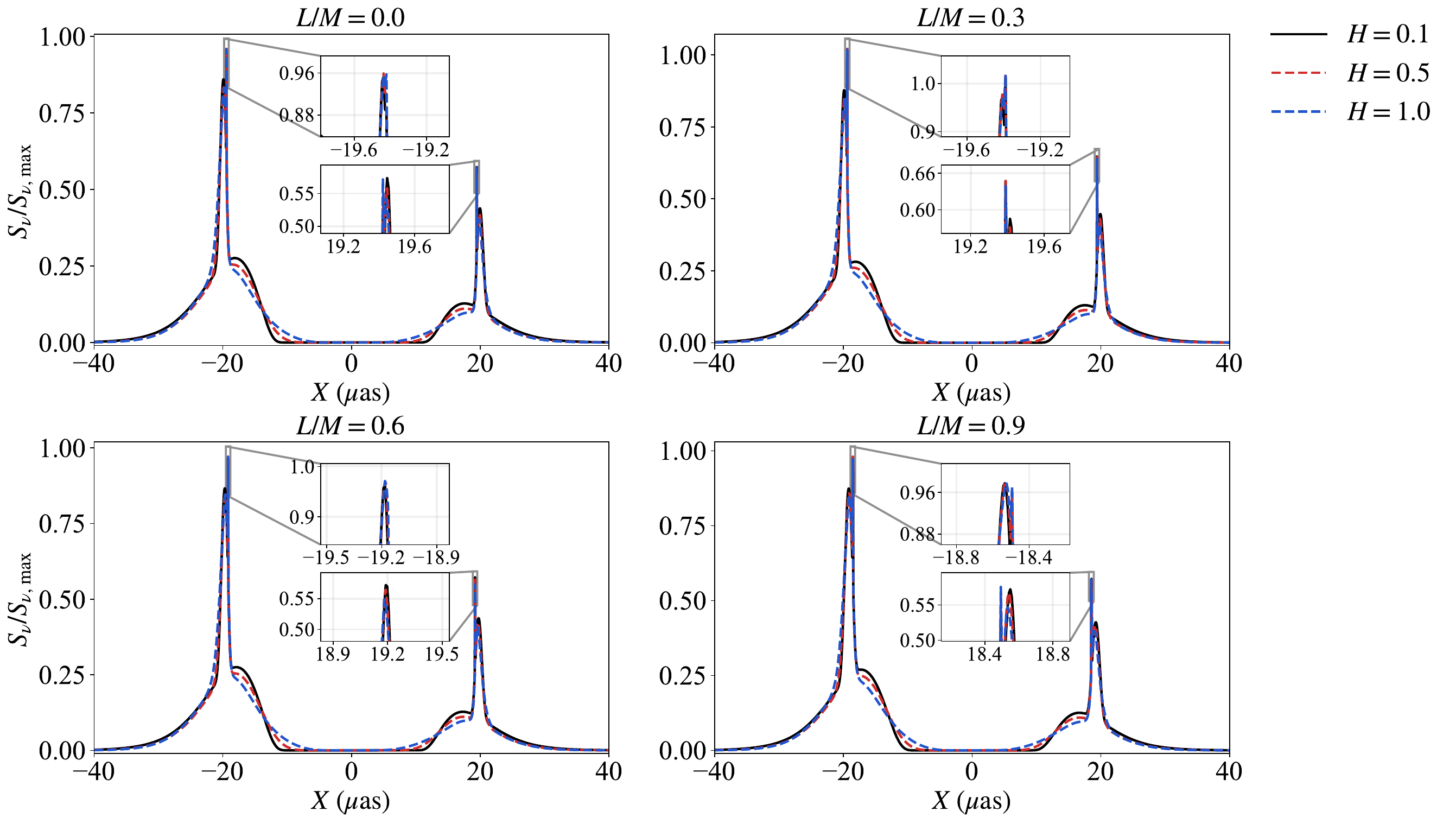}
\caption{Horizontal intensity cross-sections along $Y=0$ regrouped by fixed core scale $L/M$, highlighting how variations in disk thickness $H=0.1, 0.5, 1.0$ modify the inner ring shoulders and outer extended emission tails.}
\label{fig:RIAF_thickness_2-1}
\end{figure}

\begin{figure}[tbp]
\centering
\includegraphics[width=\textwidth]{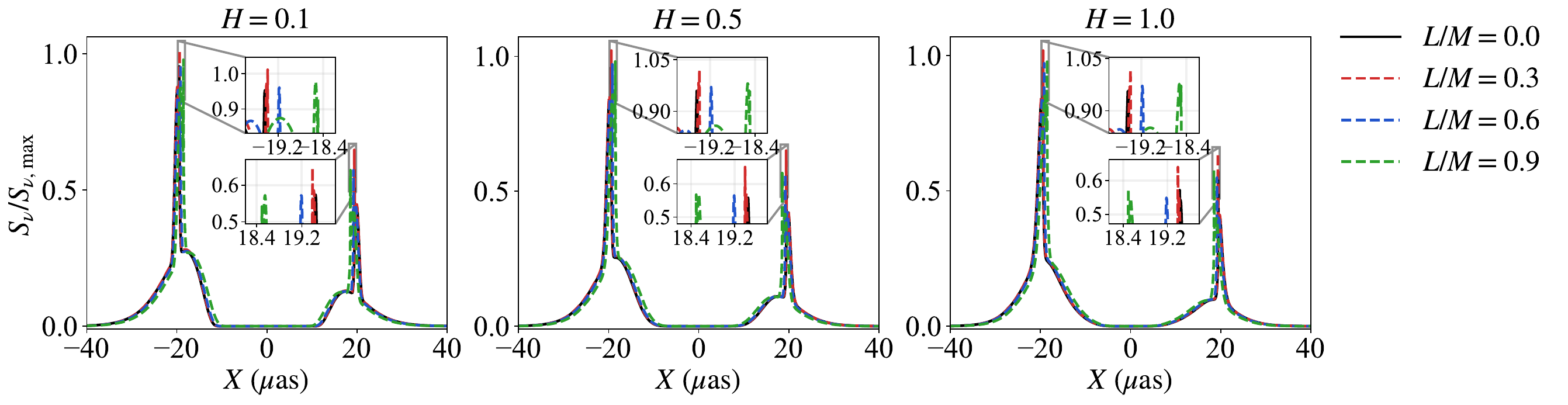}
\caption{Horizontal intensity cross-sections along $Y=0$ for the thickness sequence, comparing the different core scales $L/M = 0, 0.3, 0.6, 0.9$ at each fixed disk thickness $H=0.1, 0.5, 1.0$. Locally refined sub-pixel sampling resolves the narrow photon-ring peaks between standard image pixels.}
\label{fig:RIAF_thickness_2-2}
\end{figure}

Interestingly, under fixed-flux calibration, thickening the accretion flow primarily contracts the outer cumulative-flux envelope $R_{90}$ while leaving the half-flux radius $R_{50}$ remarkably stable. For the Schwarzschild case, as $H$ increases from $0.1$ to $1.0$, $R_{50}$ remains tightly bounded between $20.13$ and $20.22\,\mu\mathrm{as}$, whereas $R_{90}$ contracts from $27.516$ to $26.312\,\mu\mathrm{as}$. An identical trend holds at $L/M=0.9$, where $R_{50}$ stays within $19.30$--$19.38\,\mu\mathrm{as}$ across all three thicknesses while $R_{90}$ shrinks from $26.325$ to $25.179\,\mu\mathrm{as}$. Physically, although a geometrically thicker disk extends plasma to larger polar angles, the concomitant reduction in $n_{e,0}$ and the associated drop in $B$ suppresses the disk emission, thereby concentrating $90\%$ of the total flux within a smaller angular radius.

Despite this redistribution of this kind of emission, the radial imprint of the regular core scale remains highly insensitive to disk thickness. At every sampled value of $H$, increasing $L/M$ from $0$ to $0.9$ contracts $R_{50}$ by an almost identical margin of $0.82$--$0.83\,\mu\mathrm{as}$. According to Figs.~\ref{fig:RIAF_thickness_2-1} and \ref{fig:RIAF_thickness_2-2}, varying $H$ at fixed $L/M$ reshapes the inner brightness shoulders and outer radial tails of the profiles subtly, while the narrow peaks consistently shift inward. The horizontal intensity cuts in Figs.~\ref{fig:RIAF_thickness_2-1} and \ref{fig:RIAF_thickness_2-2} corroborate this robustness. Therefore, it is necessay to track the critical curve radius, $R_{50}$, and $R_{90}$ jointly in order to separate pure spacetime contraction from astrophysical scale-height effects.


\subsection{Observing Frequency}
\begin{figure}[tbp]
\centering
\includegraphics[width=\textwidth]{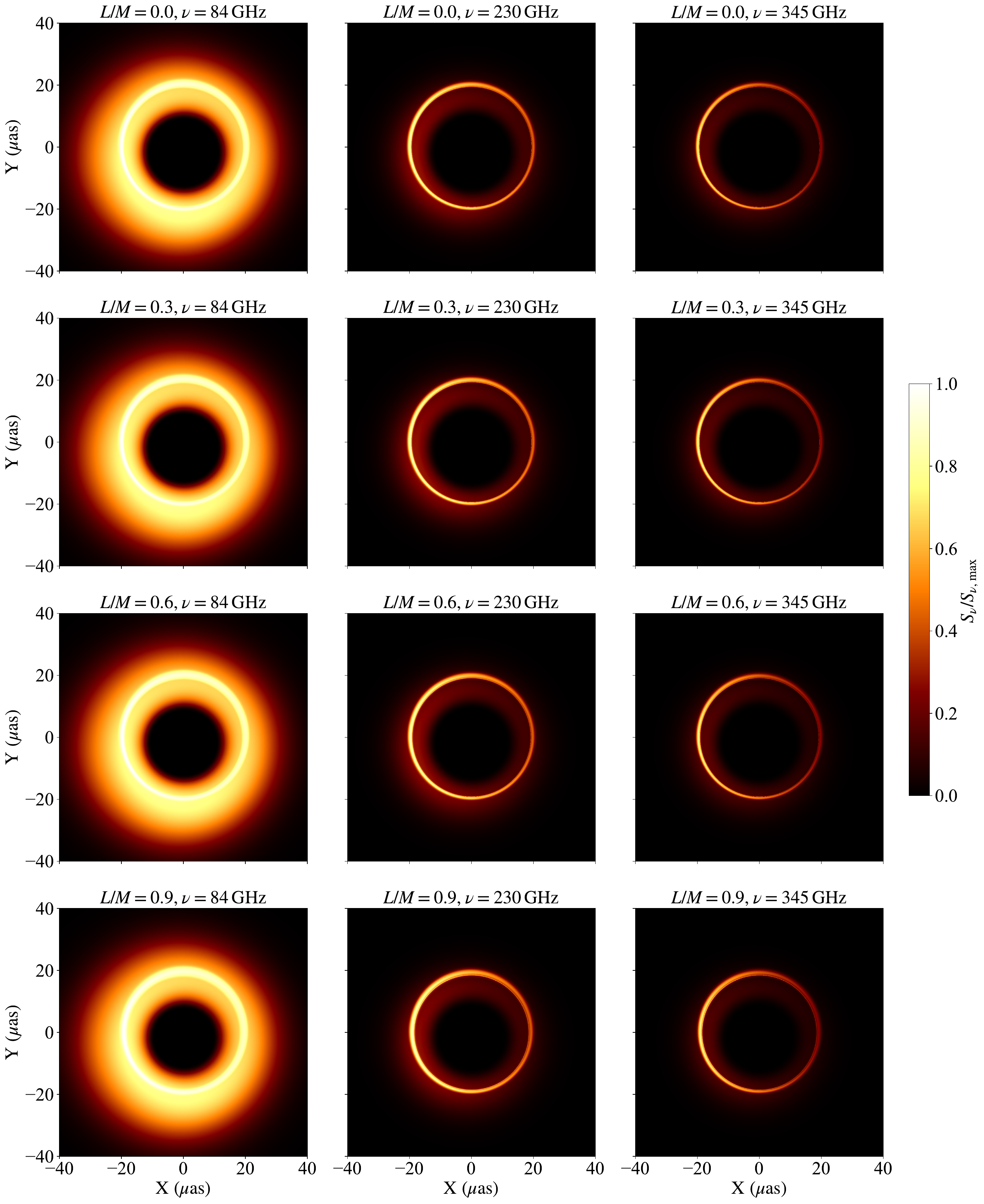}
\caption{Total-intensity synchrotron images across different observing frequencies. Columns from left to right correspond to $\nu_o = 84, 230, 345\,\mathrm{GHz}$, and rows from top to bottom correspond to $L/M=0, 0.3, 0.6, 0.9$, evaluated at $\theta=163^\circ$ and $H=0.1$.}
\label{fig:RIAF_frequency_1}
\end{figure}

Changing the observing frequency across the millimeter and submillimeter bands probes distinct optical-depth regimes and radial emission zones within the accretion flow~\cite{broderick2006frequency,lu2023a,desire2025multifrequency,wang2026imaging}. Fig.~\ref{fig:RIAF_frequency_1} compares the synthetic images across these three frequencies at $\theta=163^\circ$ and $H=0.1$. Visually, the $84\,\mathrm{GHz}$ emission forms a diffuse bright ring shifted outward from the central depression, whereas at $230$ and $345\,\mathrm{GHz}$ the emission concentrates into a sharp, narrow ring closely hugging the critical curve. Quantitatively, for $L/M=0$, increasing the frequency from $84$ to $230$ and $345\,\mathrm{GHz}$ reduces the half-flux radius $R_{50}$ from $24.763$ to $20.205$ and $19.982\,\mu\mathrm{as}$, and contracts the outer radius $R_{90}$ from $36.299$ to $27.516$ and $26.017\,\mu\mathrm{as}$, while raising the left-half flux fraction $f_{\rm left}$ from $54.06\%$ to $62.38\%$ and $64.91\%$. An analogous compaction occurs for $L/M=0.9$, where $R_{50}$ drops from $23.983$ to $19.382$ and $19.168\,\mu\mathrm{as}$, $R_{90}$ falls from $35.564$ to $26.325$ and $24.733\,\mu\mathrm{as}$. Because the spacetime geometry is identical across frequencies at fixed $L/M$, this strong chromatic evolution is driven entirely by synchrotron spectral weighting and self-absorption.

In a thermal synchrotron plasma, the radial weighting of emission is governed by the ratio of the comoving frequency $\nu_e = \nu_o/g$ to the local critical synchrotron frequency $\nu_s \propto B\Theta_e^2\sin\vartheta_B$, where $\Theta_e \equiv k_B T_e/(m_e c^2)$ is the dimensionless electron temperature~\cite{mahadevan1996harmony,dexter2016a}. Given our radial profiles in Eqs.~\eqref{eq:ne}--\eqref{eq:magnetic} with $\delta=1.1$ and $\alpha=0.84$, the magnetic field and characteristic frequency scale radially as $B \propto n_{e,0}^{1/2}r^{-1.05}$ and $\nu_s \propto n_{e,0}^{1/2}r^{-2.73}$ at fixed latitude and pitch angle. Higher observing frequencies therefore preferentially sample the hotter, more strongly magnetized innermost radii near the photon sphere. Furthermore, because the thermal synchrotron spectrum falls off exponentially at high frequencies, matching the fixed $0.5\,\mathrm{Jy}$ flux at higher $\nu_o$ requires a progressively denser plasma.

\begin{figure}[tbp]
\centering
\includegraphics[width=\textwidth]{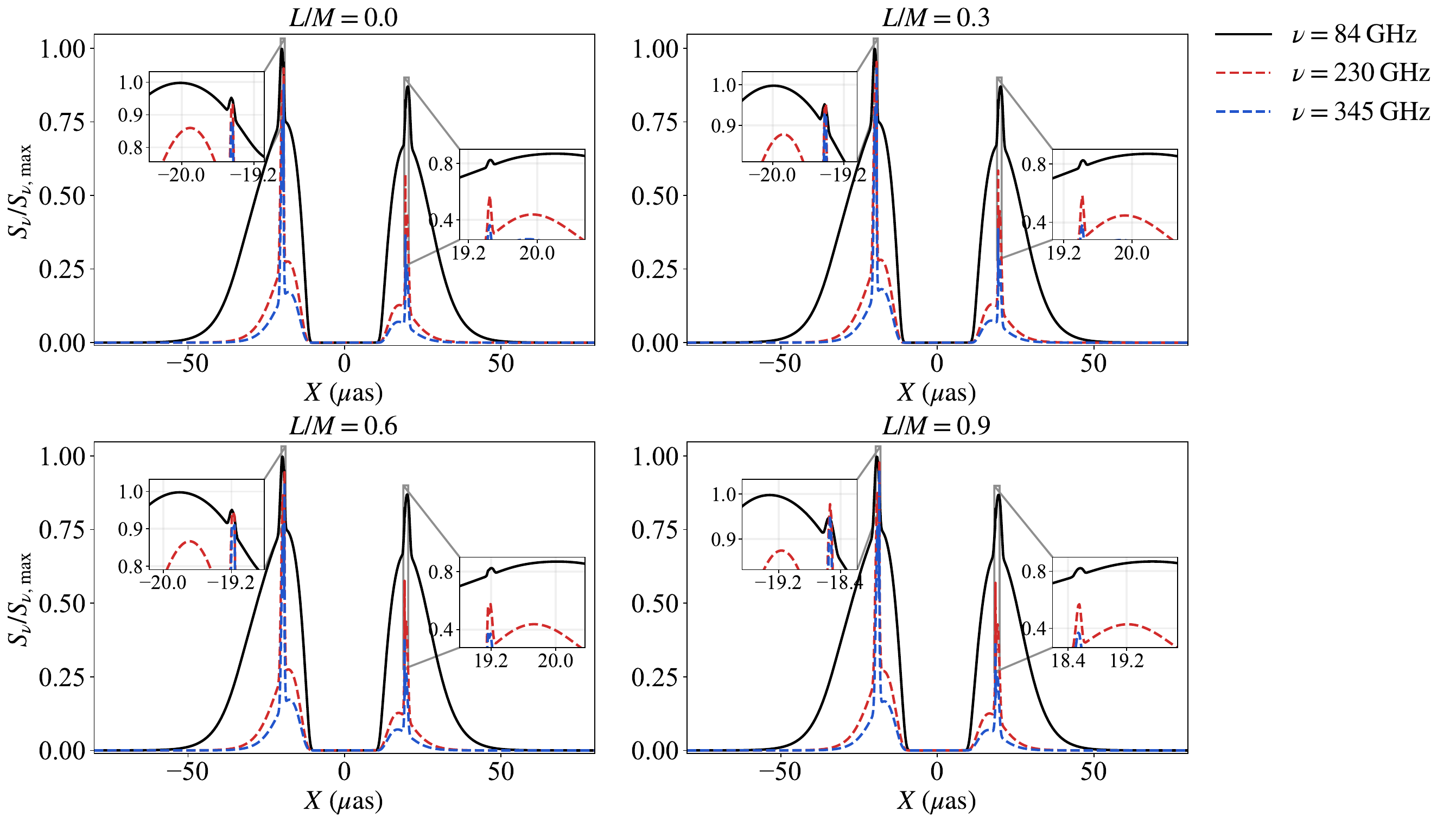}
\caption{Horizontal intensity cross-sections along $Y=0$ regrouped by fixed core scale $L/M$, illustrating the transition from an outer emission profile at $84\,\mathrm{GHz}$ to peaks near the critical curve at $230$ and $345\,\mathrm{GHz}$.}
\label{fig:RIAF_frequency_2-1}
\end{figure}

\begin{figure}[tbp]
\centering
\includegraphics[width=\textwidth]{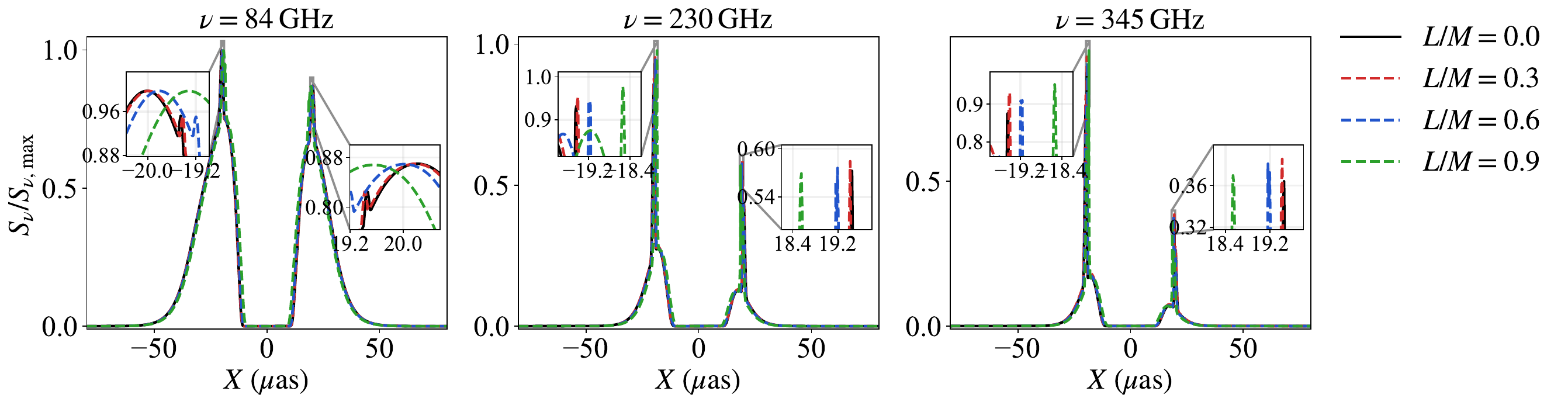}
\caption{Horizontal intensity cross-sections along $Y=0$ for the frequency sequence, comparing the four core scales $L/M = 0, 0.3, 0.6, 0.9$ at three different observing frequency $\nu_o = 84, 230, 345\,\mathrm{GHz}$. Insets highlight the inward shift of the brightness peaks with increasing $L/M$.}
\label{fig:RIAF_frequency_2-2}
\end{figure}

Despite the higher electron density normalization at $230$ and $345\,\mathrm{GHz}$, the rapid decline of the synchrotron absorption cross-section with frequency renders the accretion flow substantially more optically thin at shorter wavelengths~\cite{broderick2006frequency,desire2025multifrequency}. At $84\,\mathrm{GHz}$, strong synchrotron self-absorption obscures the innermost lensed trajectories and shifts the effective photosphere outward, diluting the relativistic Doppler asymmetry with $f_{\rm left} \approx 54\%$. Nevertheless, rays passing strictly adjacent to the critical curve can still accumulate considerable optical depth even at high frequencies due to their logarithmic orbital winding around the photon sphere~\cite{gralla2019black}. Figs.~\ref{fig:RIAF_frequency_2-1} and \ref{fig:RIAF_frequency_2-2} demonstrate that photon-ring peaks remain partially self-absorbed even when the surrounding disk is optically thin.

The horizontal cross-sections in Figs.~\ref{fig:RIAF_frequency_2-1} and \ref{fig:RIAF_frequency_2-2} clearly illustrate how self-absorption decouples the $84\,\mathrm{GHz}$ brightness peak from the geometric critical curve. At $84\,\mathrm{GHz}$, the peak-to-peak diameter of the intensity profile exceeds the geometric critical curve diameter $5.09\%$ at $L/M=0.9$, whereas at $230$ and $345\,\mathrm{GHz}$ the narrow lensed peaks closely trace the critical boundary. Moreover, the plasma density adjustment required at fixed flux becomes more pronounced at higher frequencies. Increasing $L/M$ from $0$ to $0.9$ lowers the calibrated normalization $n_{e,0}$ by $4.59\%$, $9.15\%$, and $10.38\%$ at $84$, $230$, and $345\,\mathrm{GHz}$, respectively, with the $230\,\mathrm{GHz}$ adjustment corresponding to a $4.68\%$ decrease in the magnetic field strength through $B\propto n_{e,0}^{1/2}$. Because optically thick $84\,\mathrm{GHz}$ emission originates predominantly from larger radii where the metric approaches the Schwarzschild form, its required density normalization is less sensitive to the near-horizon core scale $L/M$ than the optically thin $230$ and $345\,\mathrm{GHz}$ emission originating deep within the strong-field region.


\section{Finite-Resolution Morphology and Image Contrast}
\label{sec5}
\begin{figure}[tbp]
\centering
\includegraphics[width=\textwidth]{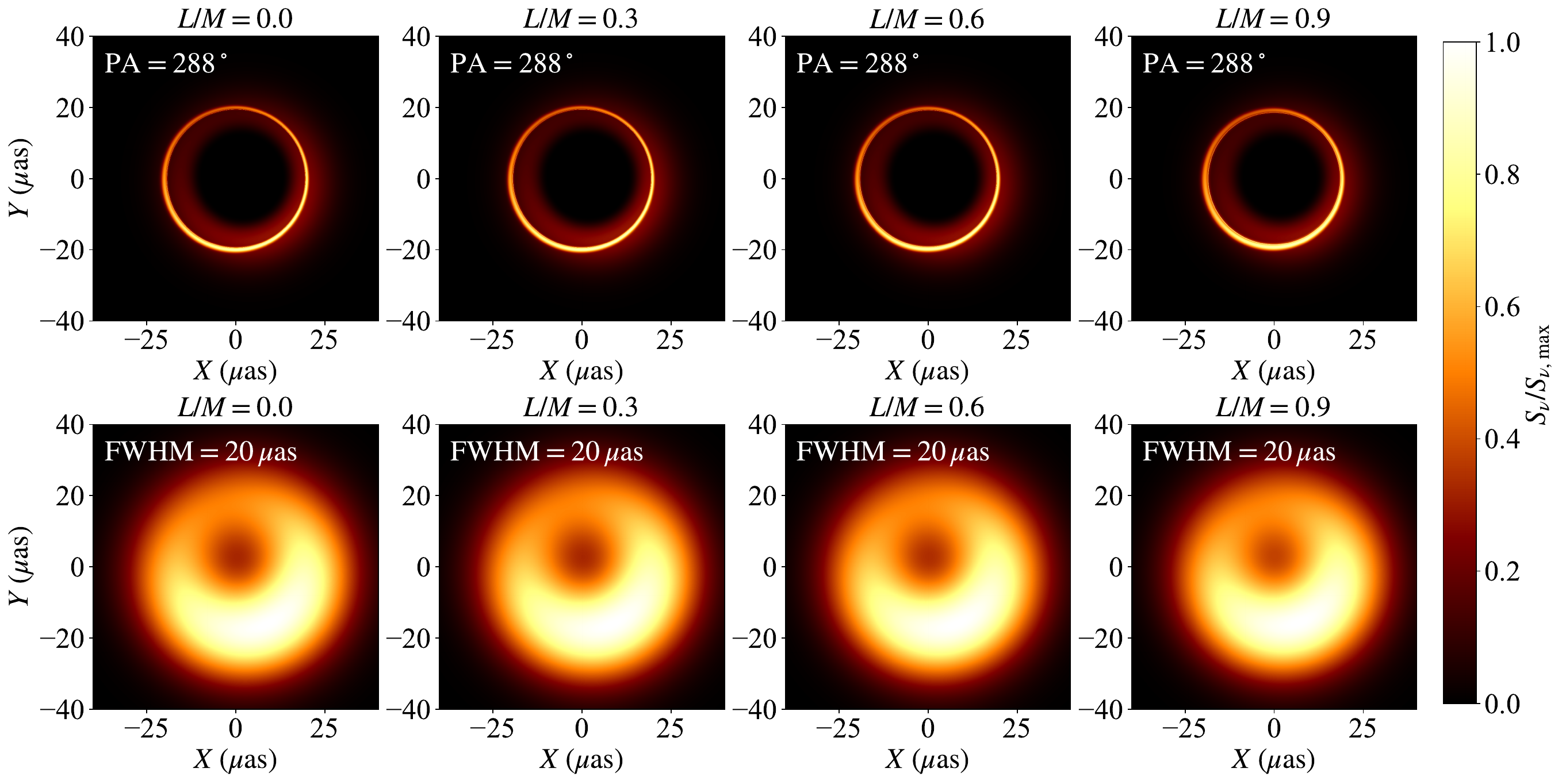}
\caption{Simulated RIAF images (top row) and the corresponding Gaussian-blurred images with a FWHM of $w=20\,\mu\mathrm{as}$ (bottom row) for different values of $L/M$, evaluated at $\theta=163^\circ$, $H=0.1$, $\nu_o=230\,\mathrm{GHz}$, and a position angle of $\mathrm{PA}=288^\circ$.}
\label{fig:RIAF_images}
\end{figure}

To facilitate a direct comparison with the EHT observations of $\mathrm{M87}^*$, Fig.~\ref{fig:RIAF_images} displays the unblurred ray-traced RIAF images (top row) alongside their beam-convolved counterparts (bottom row) across the core scales $L/M$, oriented at a viewing inclination of $\theta=163^\circ$ and a sky position angle of $\mathrm{PA}=288^\circ$~\cite{walker2018structure,cui2023precessing,chen2026images}. To imitate the finite angular resolution of current EHT instrumentation, we convolve the synthetic intensity maps with a circular Gaussian beam of full width at half maximum (FWHM) $w=20\,\mu\mathrm{as}$~\cite{akiyama2019firsta,akiyama2021firsta,akiyama2022firsta}. After $20\,\mu\mathrm{as}$ Gaussian smoothing, the synthetic images across all four core scales reproduce the characteristic asymmetric bright ring and central brightness depression observed in $\mathrm{M87}^*$.

To quantitatively assess whether these strong-field deviations can be resolved by future facilities such as the next-generation EHT (ngEHT) and the Black Hole Explorer (BHEX)~\cite{uniyal2026future,chen2026images}, we evaluate two pixel-wise image comparison metrics between each regular black hole image $I$ ($L/M > 0$) and the Schwarzschild reference image $K$ ($L/M = 0$). First, we compute the normalized cross-correlation coefficient ($\mathrm{nCCC}$), which measures the spatial overlap between two images such that $\mathrm{nCCC}=1$ for identical distributions and $\mathrm{nCCC}=0$ for completely uncorrelated distributions~\cite{mizuno2018the,gold2020verification,uniyal2026future}. The corresponding cross-correlation mismatch is given by~\cite{wang2004image,chen2026images}
\begin{align}
\label{eq:nCCC}
\mathcal{M}&\equiv1-\mathrm{nCCC}\\\notag
&=1-\dfrac{1}{N}\sum_{i=1}^{N}\dfrac{\left(I_i-\mu_I\right)\left(K_i-\mu_K\right)}{\sigma_I\sigma_K},
\end{align}
where the summation runs over all $N$ pixels in the field of view, and $\mu_I$ and $\sigma_I$ denote the mean pixel intensity and sample standard deviation of image $I$, respectively,
\begin{align}
\mu_I&=\dfrac{1}{N}\sum_{i=1}^{N} I_i,\\
\sigma_I^2&=\dfrac{1}{N-1}\sum_{j=1}^{N}(I_j-\mu_I)^2.
\end{align}
Similarly for $\mu_K$ and $\sigma_K$. Ixn addition, as a complementary diagnostic sensitive to both structural correlation and relative variance amplitude, we introduce the structural dissimilarity index, which is defined as~\cite{wang2004image,mizuno2018the,chen2026images,uniyal2026future}
\begin{align}
\mathcal{D}\equiv\left|\dfrac{\sigma_I^2+\sigma_K^2}{2\sigma_{IK}}\right|-1,
\end{align}
with the spatial sample covariance given by
\begin{align}
\sigma_{IK}=\dfrac{1}{N-1}\sum_{j=1}^{N}\left(I_j-\mu_I\right)\left(K_j-\mu_K\right).
\end{align}

\begin{figure}[tbp]
\centering
\includegraphics[width=\textwidth]{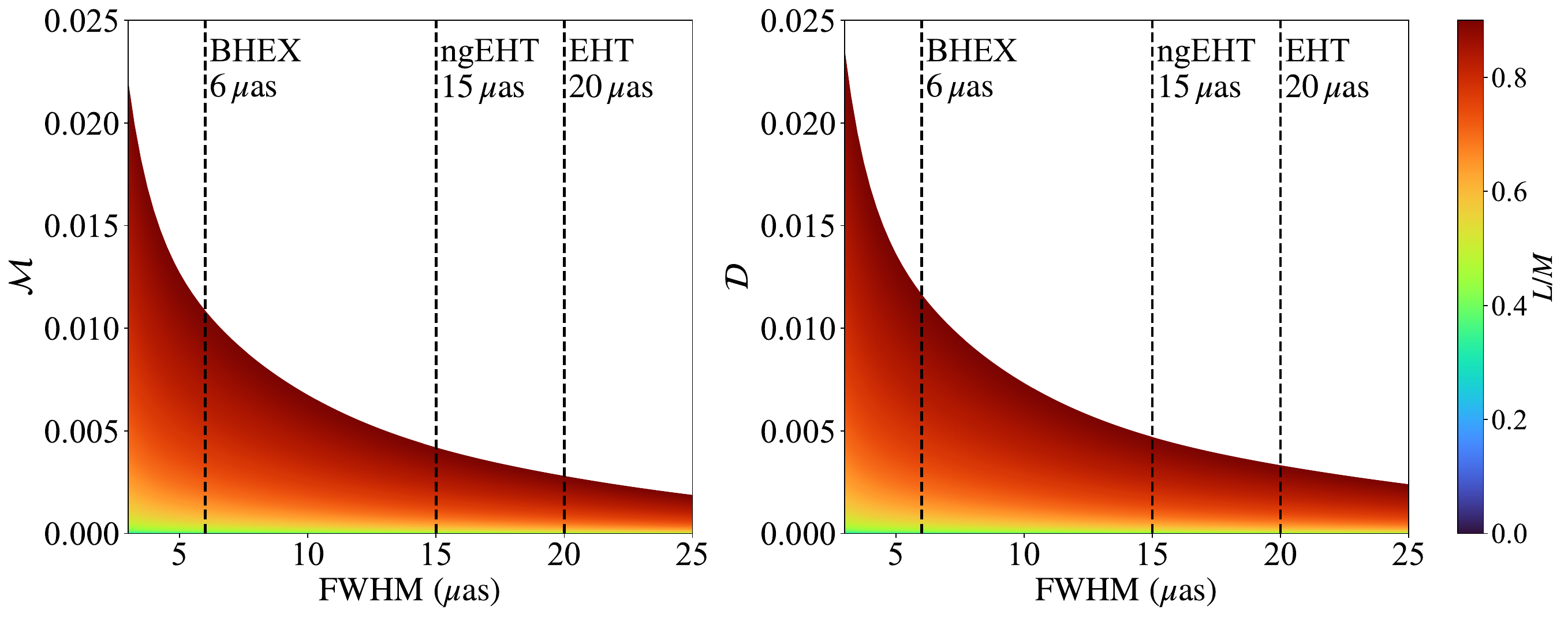}
\caption{Image-comparison metrics $\mathcal{M}=1-\mathrm{nCCC}$ (left panel) and $\mathcal{D}=\mathrm{DSSIM}$ (right panel) as a function of the FWHM. The colorbar indicates the parameter $L/M \in [0, 0.9]$, and the vertical dashed lines mark the nominal angular resolutions of BHEX ($6\,\mu\mathrm{as}$), ngEHT ($15\,\mu\mathrm{as}$), and the current EHT ($20\,\mu\mathrm{as}$).}
\label{fig:metrics}
\end{figure}

\begin{table}[tbp]
\centering
\caption{Image-comparison metrics $\mathcal{M}$ and $\mathcal{D}$ and their ratio relative to the case $L/M=0$ at the characteristic angular resolutions of BHEX ($6\,\mu\mathrm{as}$), ngEHT ($15\,\mu\mathrm{as}$), and EHT ($20\,\mu\mathrm{as}$).}
\label{tab:metrics}
\setlength{\tabcolsep}{14pt}
\renewcommand{\arraystretch}{1.2}
\begin{tabular}{ccccc}
\toprule
$L/M$ & Metric & BHEX ($6\,\mu\mathrm{as}$) & ngEHT ($15\,\mu\mathrm{as}$) & EHT ($20\,\mu\mathrm{as}$) \\
\midrule
\multirow{3}{*}{0.3}
& $\mathcal M$ & $1.41 \times 10^{-5}$ & $5.55 \times 10^{-6}$ & $3.79 \times 10^{-6}$ \\
& $\mathcal D$ & $1.48 \times 10^{-5}$ & $5.97 \times 10^{-6}$ & $4.24 \times 10^{-6}$ \\
& $\mathcal D/\mathcal M$ & $1.05$ & $1.08$ & $1.12$ \\
\cmidrule(lr){1-5}
\multirow{3}{*}{0.6}
& $\mathcal M$ & $8.66 \times 10^{-4}$ & $3.36 \times 10^{-4}$ & $2.25 \times 10^{-4}$ \\
& $\mathcal D$ & $9.16 \times 10^{-4}$ & $3.72 \times 10^{-4}$ & $2.64 \times 10^{-4}$ \\
& $\mathcal D/\mathcal M$ & $1.06$ & $1.11$ & $1.17$ \\
\cmidrule(lr){1-5}
\multirow{3}{*}{0.9}
& $\mathcal M$ & $1.09 \times 10^{-2}$ & $4.21 \times 10^{-3}$ & $2.82 \times 10^{-3}$ \\
& $\mathcal D$ & $1.17 \times 10^{-2}$ & $4.71 \times 10^{-3}$ & $3.34 \times 10^{-3}$ \\
& $\mathcal D/\mathcal M$ & $1.07$ & $1.12$ & $1.18$ \\
\bottomrule
\end{tabular}
\end{table}

Fig.~\ref{fig:metrics} and Tab.~\ref{tab:metrics} present the evolution of the cross-correlation mismatch $\mathcal{M}$ and the structural dissimilarity $\mathcal{D}$ as functions of the Gaussian beam FWHM across the continuous core-scale spectrum $L/M \in [0, 0.9]$, with vertical dashed lines marking the characteristic angular resolutions of BHEX ($6\,\mu\mathrm{as}$), ngEHT ($15\,\mu\mathrm{as}$), and EHT ($20\,\mu\mathrm{as}$)~\cite{uniyal2026future,chen2026images}. Both $\mathcal{M}$ and $\mathcal{D}$ decrease monotonically with increasing beam size, reflecting the progressive blending of fine ring structures at lower angular resolution. At any fixed beam width, the image discrepancy grows nonlinearly with the core scale $L/M$ due to the accelerated contraction of the photon sphere and emission ring near $L/M = 0.9$. Specifically, as shown in Tab.~\ref{tab:metrics}, improving the angular resolution from the current EHT baseline to ngEHT and BHEX could increase $\mathcal{M}$ from $2.82\times 10^{-3}$ to $4.21\times 10^{-3}$ and $1.09\times 10^{-2}$ for $L/M=0.9$, respectively, while $\mathcal{D}$ rises from $3.34\times 10^{-3}$ to $4.71\times 10^{-3}$ and $1.17\times 10^{-2}$. By comparison, the values are $\mathcal{M}=2.20\times 10^{-2}$ and $\mathcal{D}=2.35\times 10^{-2}$ for $3\,\mu\mathrm{as}$. Furthermore, $\mathcal{D}$ in our models consistently exceeds the cross-correlation mismatch $\mathcal{M}$, maintaining a ratio of $\mathcal{D}/\mathcal{M} \simeq 1.05\text{--}1.18$ across the three benchmark facilities. Nevertheless, the fact that both metrics remain on the order of $10^{-3}$ at $15\text{--}20\,\mu\mathrm{as}$, reaching the $10^{-2}$ level only for large parameter $L/M\simeq0.9$ at $6\,\mu\mathrm{as}$, demonstrates that distinguishing regular black holes with hedgehog scalar hair from a standard Schwarzschild black hole via total-intensity images will require space-VLBI angular resolutions comparable to or finer than BHEX~\cite{uniyal2026future,chen2026images}.


\section{Conclusion}
\label{sec6}
In this work, we have systematically investigated the millimeter and submillimeter radiative signatures of geometrically regular black holes endowed with hedgehog scalar hair~\cite{bahamonde2026geometrically,nashed2026odd}, illuminated by the stationary and axisymmetric RIAF calibrated to the observational parameters of $\mathrm{M87}^*$. By fixing total flux density of $0.5\,\mathrm{Jy}$ across all configurations, our analysis disentangles the interplay in this spacetime between strong-field geometric contraction, plasma density recalibration, chromatic self-absorption, and finite instrumental resolution.

The most robust signature of the regular core scale $L/M$ across our fixed-flux models is a systematic radial contraction of the emission structure. At a fixed mass $M$, increasing the core scale $L$ shrinks the photon sphere and critical impact parameter. At $230\,\mathrm{GHz}$, as $L/M$ increases from $0$ to $0.9$, the half-flux radius $R_{50}$ contracts from $20.205$ to $19.382\,\mu\mathrm{as}$ in the fiducial $\theta=163^\circ, H=0.1$ setup,. Indeed, the dimensionless ratio $R_{50}/b_p$ varies by only $\sim 0.75\%$, confirming that cumulative-flux radii preserve a tight, calibratable correspondence with the underlying critical curve across diverse viewing inclinations and disk thicknesses.

Comparing competing spacetime geometries at an identical observed flux density makes plasma recalibration an integral part of the physical interpretation~\cite{akiyama2019firstb,akiyama2022firstb}. This coupling between geometry and plasma state is also evident in the disk scale-height scan. Fixed-flux image comparisons therefore map out a self-consistent trajectory in plasma parameter space rather than a purely geometric projection. Furthermore, multi-frequency synthesis at $84$, $230$, and $345\,\mathrm{GHz}$ demonstrates how synchrotron spectral weighting and self-absorption modulate the observability of the strong-field region~\cite{broderick2006frequency,lu2023a,desire2025multifrequency}. Because both electron temperature and magnetic field strength increase inward, higher observing frequencies preferentially sample the innermost lensed trajectories near the photon sphere and exhibit substantially lower optical depths. Higher-frequency submillimeter observations thus provide a significantly cleaner probe of near-horizon regular-core modifications.

Finally, convolving the synthetic images with Gaussian beams highlights the challenge of resolving regular-core deviations under current baseline capabilities. Increasing the beam FWHM from $3$ to $20\,\mu\mathrm{as}$ mixes ring flux into the central depression and suppresses both the cross-correlation mismatch $\mathcal{M}$ and structural dissimilarity $\mathcal{D}$ by more than $85\%$. Because these mismatches lie on the order of $10^{-3}$ or below at $20\,\mu\mathrm{as}$, resolving the imprint of hedgehog scalar hair requires the finer angular resolution of future space-VLBI missions such as BHEX~\cite{uniyal2026future,chen2026images}. Since the fundamental length scale $L = \eta/\mu_\star$ is fixed by action parameters in Bahamonde's theory~\cite{bahamonde2026geometrically}, our $L/M$ scan at fixed mass directly tests candidate theoretical scales. Future extensions incorporating full linear and circular polarization structures~\cite{akiyama2021firsta,akiyama2021firstb,yin2026circular,feng2026polarization}, joint multifrequency fitting with a unified plasma density normalization, and direct comparisons in the interferometric visibility and closure-quantity domains~\cite{chael2018interferometric,blackburn2020closure,johnson2020universal} will further sharpen horizon-scale tests of regular black-hole geometries.


\section*{Acknowledgments}

This work was supported in part by NSFC under Grants No. 12375042 and No. 11975046.


\bibliography{reference}

\end{document}